\newcommand{\NDMAP}{Ni(C$_5$D$_{14}$N$_2$)$_2$N$_3$(PF$_6$)}

\newcommand{\NDMAZ}{Ni(C$_5$H$_{14}$N$_2$)$_2$N$_3$(ClO$_4$)}
\newcommand{\NTENP}{[Ni($N,N'$-bis(3-aminopropyl)propane-1,3-diamine($\mu$-NO$_2$)]ClO$_4$}

\documentclass[aps,prb,twocolumn,superscriptaddress,showpacs,,amsmath,amssymb]{revtex4}
\usepackage{graphicx}
\usepackage{natbib}
\usepackage{bm}

\begin{document}

\title{Half-ordered state in the anisotropic Haldane-gap antiferromagnet \NDMAP.}

\author{A. Zheludev}
\affiliation{Condensed Matter Sciences Division, Oak Ridge
National Laboratory, Oak Ridge, TN 37831-6393, USA.}
 \email{zheludevai@ornl.gov}
 \homepage{http://neutron.ornl.gov/~zhelud/}

\author{B. Grenier}
\author{E. Ressouche}
\author{L.-P.~Regnault}
\affiliation{DRFMC/SPSMS/MDN, CEA-Grenoble, 17 rue des Martyrs,
38054 Grenoble Cedex, France.}

\author{Z. Honda}
\affiliation{Faculty of Engineering, Saitama University, Urawa,
Saitama 338-8570, Japan.}

\author{K. Katsumata}
\affiliation{The RIKEN Harima Institute, Mikazuki, Sayo, Hyogo
679-5148, Japan.}

\date{\today}
\begin{abstract}
Neutron diffraction experiments performed on the Haldane gap
material\ \NDMAP  in high magnetic fields applied at an angle to
the principal anisotropy axes reveal two consecutive field-induced
phase transitions. The low-field phase is the gapped Haldane
state, while at high fields the system exhibits 3-dimensional
long-range N\'eel order. In a peculiar phase found at intermediate
fields only half of all the spin chains participate in the
long-range ordering, while the other half remains disordered and
gapped.

\end{abstract}


\maketitle

High-field phase transitions in quantum antiferromagnets (AFs)
have recently drawn a great deal of attention. Of particular
interest is field-induced spin freezing exhibited by many quantum
spin liquids. The massive triplet of low-lying $S=1$ gap
excitations (magnons) in such systems is subject to Zeeman
splitting by external magnetic fields. A soft-mode quantum phase
transition occurs when the gap for one member of the triplet is
driven to zero. The result is a Bose condensation of magnons. In
the presence of magnetic anisotropy and weak inter-chain
interactions, always found in real materials, the magnetized
high-field phase is a N\'eel-like state with AF long-range order.
The phase transition itself and the peculiarities of the
high-field phase have been studied experimentally in several model
materials including the Haldane-gap antiferromagnets NDMAP
(\NDMAP)
\cite{Honda98,Honda99,Chen01,Zheludev2002,Zheludev2003,Hagiwara2003}
and NDMAZ (\NDMAZ),\cite{Honda97,Zheludev2001-2} the
bond-alternating $S=1$ chain NTENP (\NTENP) \cite{Narumi2001} and
dimer systems such as TlCuCl$_3$\cite{Ruegg2003} and
Cs$_3$Cr$_2$Br$_9$.\cite{Grenier2004}

\begin{figure}
\includegraphics[width=3.3in]{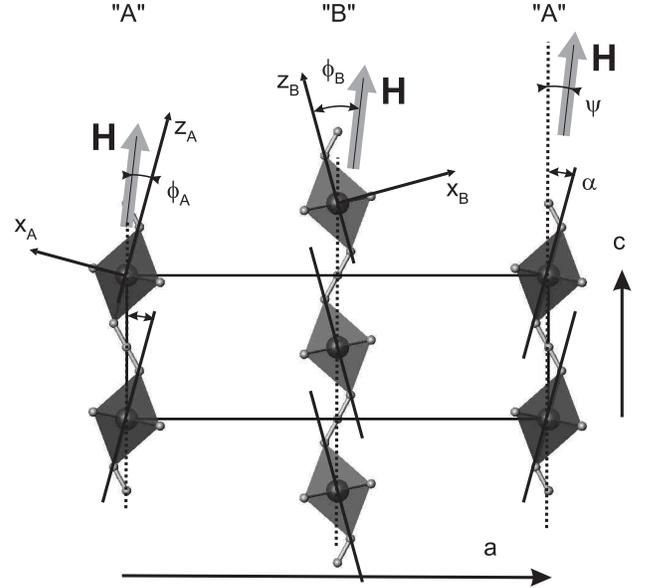}
 \caption{\label{str} Crystal structure of NDMAP in projection onto
 the $(ac)$ crystallographic  plane and the geometry of the present experiment.
 Only the N- (small balls) and Ni-ions (large balls) are shown.
 The local magnetic anisotropy axes $x$ and $z$ are tilted in the $(ac)$ plane by
 $\alpha=16^{\circ}$ relative to the $c$ axis. The tilt direction is opposite
 for type-A and type-B chains. The field $H$ is applied at an angle $\psi$ relative
 to the $c$ axis of the crystal and forms unequal angles $\phi_\mathrm{A}$ and $\phi_\mathrm{B}$ with
 the anisotropy axes $z_\mathrm{A}$ and $z_\mathrm{B}$, respectively.\label{struc}}
\end{figure}

It has been long established that magnetic anisotropy, which is
particularly important for $S=1$ materials, has a strong impact on
the phase transition. The value of the critical field $H_c$
depends on the relative orientation of the applied field and the
anisotropy tensor.\cite{Tsvelik90,Affleck90,Mitra94} A very
interesting case is that of NDMAP. This compound features two
equivalent sets of Haldane spin chains with non-collinear local
anisotropy axes. To date, all experiments were performed in
magnetic fields applied along the principal axes of the
orthorhombic crystal structure that are also the {\it macroscopic}
magnetic anisotropy axes. In these geometries all tilts of the
local anisotropy axes relative to the field direction are the same
for the two chain types. In the present study we investigate a
less symmetric scenario, in which the magnetic field is applied in
a general direction relative to the crystal axes. We find {\it
two} consecutive field-induced transitions from the
quantum-disordered spin liquid to ordered N\'eel phase, with a
novel ``half-ordered'' phase in-between.

The crystal structure of NDMAP (orthorhombic space group $P
{nmn}$, $a=18.046$~\AA, $b=8.705$~\AA, and $c=6.139$~\AA) was
described in detail in Ref.~\cite{Zheludev2003}. The $S=1$ chains
run along the crystallographic $c$ axis. The in-chain exchange
constant is $J=2.6$~meV. Inter-chain interactions are much weaker:
$|J_\bot/J|<10^{-3}$. Magnetic anisotropy is predominantly of
single-ion easy-plane type with $D/J\approx 0.25$. In addition,
there is a weak in-plane anisotropy term, and the degeneracy of
Haldane triplet is fully lifted. The gap energies for excitations
polarized along the principal anisotropy axes $x$, $y$ and $z$ are
$\Delta_x=0.42(3)$~meV, $\Delta_y=0.52(6)$~meV, and
$\Delta_z=1.9(1)$~meV.\cite{Zheludev2001} The anisotropy axes are
determined by the geometry of the corresponding Ni$^{2+}$
coordination octahedra and, as mentioned above, do {\it not}
exactly coincide with crystallographic directions. Instead, the
$x$ and $z$ axes of the NiN$_6$ octahedra are in the $(a,c)$
crystallographic plane, but tilted by $\alpha\approx 16^\circ$
relative to the $a$ and $c$ axes, respectively. There are two
types of chains related by symmetry, and the corresponding tilt
directions are opposite. Within each set of chains the
Ni$^{2+}$-sites form a simple orthorhombic Bravais lattice. On the
other hand, the two sets of chains are displaced by $(0.5, 0.5,
0.5)$ relative to each other. The overall lattice of Ni$^{2+}$
ions is thus a body-centered one. Due to this geometric
frustration, the two sets of antiferromagnetic spin chains are
magnetically decoupled at the Mean Field level. The main features
of the crystal structure of NDMAP are illustrated in
Fig.~\ref{struc}.

Our new neutron diffraction experiments were performed on a fully
deuterated single crystal NDMAP sample of approximate mass 100~mg.
The data were take on the D23 lifting-counter diffractometer at
Institut Laue-Langevin. Sample environment was a vertical-field
cryomagnet with a dilution refrigerator insert. The data were
taken at $T\sim 35$~mK in magnetic fields up to 6~T.
Unfortunately, there was no possibility to rotate the sample {\it
in situ} during the experiment. Re-mounting the sample to explore
several orientations was not an option either, as the crystals are
known to shatter and deteriorate rapidly during   cooling/heating
cycles. For this reason only one experimental geometry was
realized, with the magnetic field applied at an angle
$\psi=14.2^\circ$ to the $c$ axis, in the $(a,c)$ crystallographic
plane, as shown in Fig.~\ref{struc}. For one set of spin chains
that we will refer to as ``type A'', the field was thus almost
exactly parallel to the main anisotropy axis, the corresponding
angle being $\phi_\mathrm{B}=\alpha-\psi=1.8^\circ$. For type-B
chains the angle between the field direction and the local
Ni$^{2+}$ anisotropy axes was considerably larger,
$\phi_\mathrm{B}=\alpha+\psi=30.2^{\circ}$.

\begin{figure}
\includegraphics[width=3.3in]{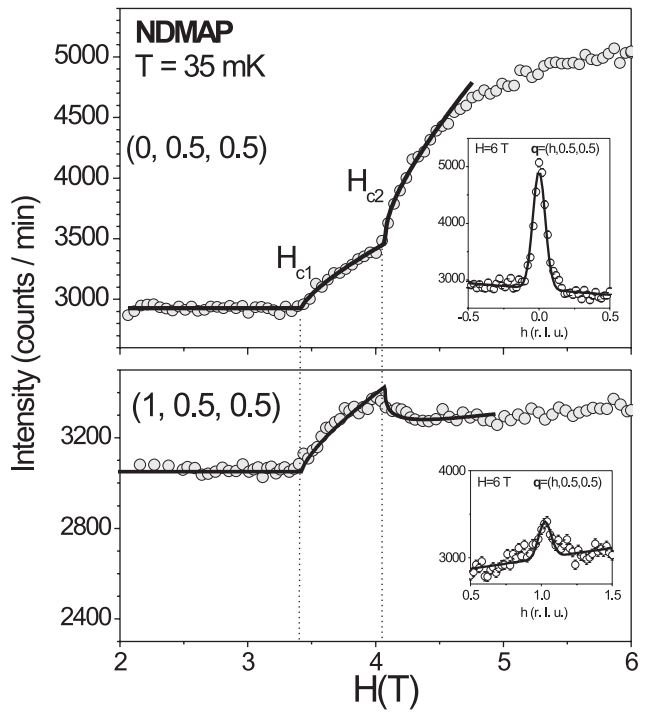}
 \caption{\label{vsH} Main panels: field dependence of two magnetic Bragg peak intensities measured in NDMAP at $T=35$~mK.
 The magnetic field is applied in the $(a,c)$ crystallographic plane at an angle 14.2$^\circ$ to the $c$ axis.
 The solid lines are as described in the text. Insets: $h$-scans across the two magnetic Bragg peaks at $H=6$~T.
 \label{data}}
\end{figure}

Our main experimental result is drawn from the measured field
dependence of the $(0,0.5,0.5)$ and $(1,0.5,0.5)$ magnetic Bragg
intensities at $T=35$~mK. These data are plotted Fig.~\ref{data}.
In the geometry of the present experiment {\it two} distinct
anomalies are detected at $H_\mathrm{c}^{(1)}=3.4$~T and
$H_\mathrm{c}^{(2)}=4.1$~T. Below $H_\mathrm{c}^{(1)}$ there is no
antiferromagnetic Bragg scattering in NDMAP that retains its
spin-singlet ground state. At $H_\mathrm{c}^{(1)}$
antiferromagnetic Bragg reflections simultaneously appear at both
$(0,0.5,0.5)$ and $(1,0.5,0.5)$ reciprocal-space positions. These
two peaks remain of roughly equal intensity in the field range
$H_\mathrm{c}^{(1)}<H<H_\mathrm{c}^{(2)}$. Above the 2nd
transition at $H_\mathrm{c}^{(2)}$ the intensity of the
$(0,0.5,0.5)$ peak starts to increase more rapidly. In contrast,
the $(1,0.5,0.5)$ peak intensity flattens out and even decreases
slightly. Typical scan across the two magnetic Bragg reflection
collected at $H=6$~T are shown in the insets in Fig.~\ref{data}.

Though a detailed determination of the magnetic structure has not
been performed, a survey of several magnetic Bragg peaks at
$H=6~T$ revealed a consistent intensity pattern. It was found that
$(h+k+l)$-even reflections are considerably stronger than the
$(h+k+l)$-odd ones. Within each of these reflection classes the
intensity is a smooth function of wave vector transfer, typical of
the combined effects of the magnetic form factor and neutron
polarization factors. Such behavior is consistent with an
almost-collinear antiferromagnetic spin arrangement on a
body-centered lattice, which was previously shown to be realized
in NDMAP in high magnetic fields applied strictly along the
high-symmetry crystallographic
directions.\cite{Chen01,Zheludev2003}

The field-dependent behavior observed in the present study is in
stark contrast that previously seen for $H\|c$ (i.e.,
$\phi_\mathrm{A}=\phi_\mathrm{B}$), where only a single transition
was detected. The unusual two-stage transition can be understood
in the framework of a simple model first proposed by I.
Harada.\cite{Harada} The key idea is that long-range ordering
occurs in each set of spin chains independently.

The critical fields for an individual spin chain in an arbitrarily
directed magnetic field can be estimated using perturbation
theory.\cite{Golinelli93,Regnault94} Though clearly
oversimplified, for a magnetic field applied parallel to any of
the principal anisotropy axes this method is known to give the
same values of $H_c$ as more sophisticated calculations based on
the quantum non-linear sigma-model\cite{Mitra94} or mapping to
Mayorana fermions.\cite{Tsvelik90} For a field in the $(x,z)$
plane applied at an angle $\phi$ to the magnetic easy axis $z$,
the perturbative result for $H_c$ is:\cite{Chen01}
 \begin{equation}
 \mu_\mathrm{B}H_\mathrm{c}=\sqrt{ \frac{\Delta_x\Delta_y\Delta_z}{g_x^2\Delta_x\sin^2\phi+g_z^2\Delta_z\cos^2\phi}}.
 \end{equation}
In this formula $g_z$ and $g_x$ are components of the Ni$^{2+}$
gyromagnetic tensor. Making use of the previously measured
gyromagnetic ratios,\cite{Honda98} for the geometry of the present
experiment one gets $H_c=3.8$~T and $H_c=4.3$~T for type-A and
type-B chains, respectively. While somewhat larger than the
measured values, these two fields can be associated with the two
observed ordering transitions at $H_\mathrm{c}^{(1)}$ and
$H_\mathrm{c}^{(2)}$.

Below $H_\mathrm{c}^{(1)}$ both types of spin chains are in a
quantum-disordered gapped state. As the external field exceeds
$H_\mathrm{c}^{(1)}$ at zero temperature, individual type-A chains
acquire long-range N\'eel order. Weak interactions between type-A
chains stabilize this ordered state at non-zero temperatures and
ensure the existence of true 3-dimensional long-range static
antiferromagnetic spin correlations. Nevertheless, considering
that inter-chain interactions are very weak, in the field range
$H_\mathrm{c}^{(1)}<H<H_\mathrm{c}^{(2)}$ type-B chains must
remain in the quantum-disordered gapped phase. The corresponding
Ni$^{2+}$ ions carry no static magnetization and do {\it not}
participate in the long-range N\'eel order. This peculiar phase of
NDMAP, where half the Haldane spin chains remain gapped while the
other half participate in long-range antiferromagnetic order can
be described as ``half-ordered'' state. In this regime the only
magnetized ions in NDMAP are located on type-A chains and form a
simple Bravais lattice. As a consequence, $(0,0.5,0.5)$ and
$(1,0.5,0.5)$ magnetic Bragg reflections have the same structure
factor. Their intensities should differ only slightly due to
slightly different form- and polarization-factors. These
intensities are proportional to the square of the staggered
magnetization on the A-sublattice: $I_{(0,0.5,0.5)} \propto
I_{(1,0.5,0.5)} \propto |m_\mathrm{A}|^2$.

The situation changes at $H_\mathrm{c}^{(2)}$ when the gap in
type-B chains closes as well, and they too acquire static
long-range AF spin correlations. Now static magnetic moments are
located on both A- and B-sublattices and form a body-centered
structure. In spite of geometric frustration, a definitive
relative alignment between spins on the two sublattices is
established via dipolar interactions and/or order-from-disorder
fluctuation effects. The two magnetic Bragg peaks are then no
longer equivalent. Assuming a collinear alignment of A- and B-type
spins, their intensities are given by $I_{(0,0.5,0.5)} \propto
|m_\mathrm{A}+m_\mathrm{B}|^2$ and $I_{(1,0.5,0.5)} \propto
|m_\mathrm{A}-m_\mathrm{B}|^2$. As both staggered magnetizations
increase with field, $I_{(0,0.5,0.5)}$ increases rapidly and
$I_{(1,0.5,0.5)}$ levels off.

To construct a phenomenalogical semi-quantitative description we
can assume that $m_\mathrm{A}=
m_\mathrm{A}^{(0)}|H/H_\mathrm{c}^{(1)}-1|^\beta\theta(H-H_\mathrm{c}^{(1)})$
and $m_\mathrm{B}= m^{(0)}_\mathrm{B}
|H/H_\mathrm{c}^{(2)}-1|^\beta\theta(H-H_\mathrm{c}^{(2)})$.
Assuming that the ordered moments on the two chain subsets are
collinear (and thus the corresponding polarization factors for
neutron scattering intensities are equal), the expression for the
measured Bragg intensities can be written as:
 \begin{widetext}
 \begin{subequations}
 \begin{equation}
 I_{(0, 0.5,0.5)}  \propto
   \left[{m^{(0)}_\mathrm{A}}|H/H_\mathrm{c}^{(1)}-1|^\beta\theta(H-H_\mathrm{c}^{(1)})
   +
   {m^{(0)}_\mathrm{B}}|H/H_\mathrm{c}^{(2)}-1|^\beta\theta(H-H_\mathrm{c}^{(2)})\right]^2,
 \end{equation}
 \begin{equation}
 I_{(1, 0.5,0.5)}  \propto
   \left[{m^{(0)}_\mathrm{A}}|H/H_\mathrm{c}^{(1)}-1|^\beta\theta(H-H_\mathrm{c}^{(1)})
   -
   {m^{(0)}_\mathrm{B}}|H/H_\mathrm{c}^{(2)}-1|^\beta\theta(H-H_\mathrm{c}^{(2)})\right]^2.
 \end{equation}
 \end{subequations}
\end{widetext}
Fitting this form to the experimental data (solid lines in
Fig.~\ref{data}) yields the following parameters:
$H_c^{(1)}=3.42(1)$~T, $H_c^{(2)}=4.07(1)$~T and $\beta=0.37(2)$
and $m^{(0)}_\mathrm{A}/m^{(0)}_\mathrm{B}=1.50(4)$. Overall, the
experimentally measured field dependencies are rather well
reproduced by the simple model.

The experimental observations of a two-stage transition and
half-ordered state in NDMAP highlight several important aspects of
field-induced spin freezing in highly anisotropic gap systems.
First, the two-stage transition is a dramatic demonstration of the
fact that gap energies and transition fields are influenced by
{\it local} anisotropy parameters, rather than macroscopic
magnetic anisotropy routinely measured with bulk methods. The
second conclusion is that the principle driving force of the phase
transition is the tendency of individual spin chains to form
long-range N\'eel order at $T=0$ in sufficiently strong fields.
Inter-chain interactions are of course needed to stabilize this
order in 3 dimensions at $T>0$, but play only a minor role in
determining the actual transition field at $T\rightarrow 0$. In
future experiments it would be very interesting to investigate the
angle-dependence of the transition fields in NDMAP in more detail.

We would like to thank Prof. Isao Harada at Okayama University for
inspiring this experimental study and Dr. S. M. Shapiro at
Brookhaven National Laboratory for his important contributions at
an early stage of the project. Work at ORNL was supported by the
U. S. Department of Energy under Contract No. DE-AC05-00OR22725.
Work at RIKEN was supported in part by a Grant-in-Aid for
Scientific Research from the Japan Society for the Promotion of
Science.


\end{document}